\documentclass{moriond}

\def\Journal#1#2#3#4{{#1} {\bf #2}, #3 (#4)}

\def\PRL{\em Phys. Rev. Lett.}

\def\be{\begin{equation}}
\def\ee{\end{equation}}
\def\bea{\begin{eqnarray}}
\def\eea{\end{eqnarray}}

\def\pt{p_\mathrm{T}}

\newcommand{\Photo}{\includegraphics[height=35mm]{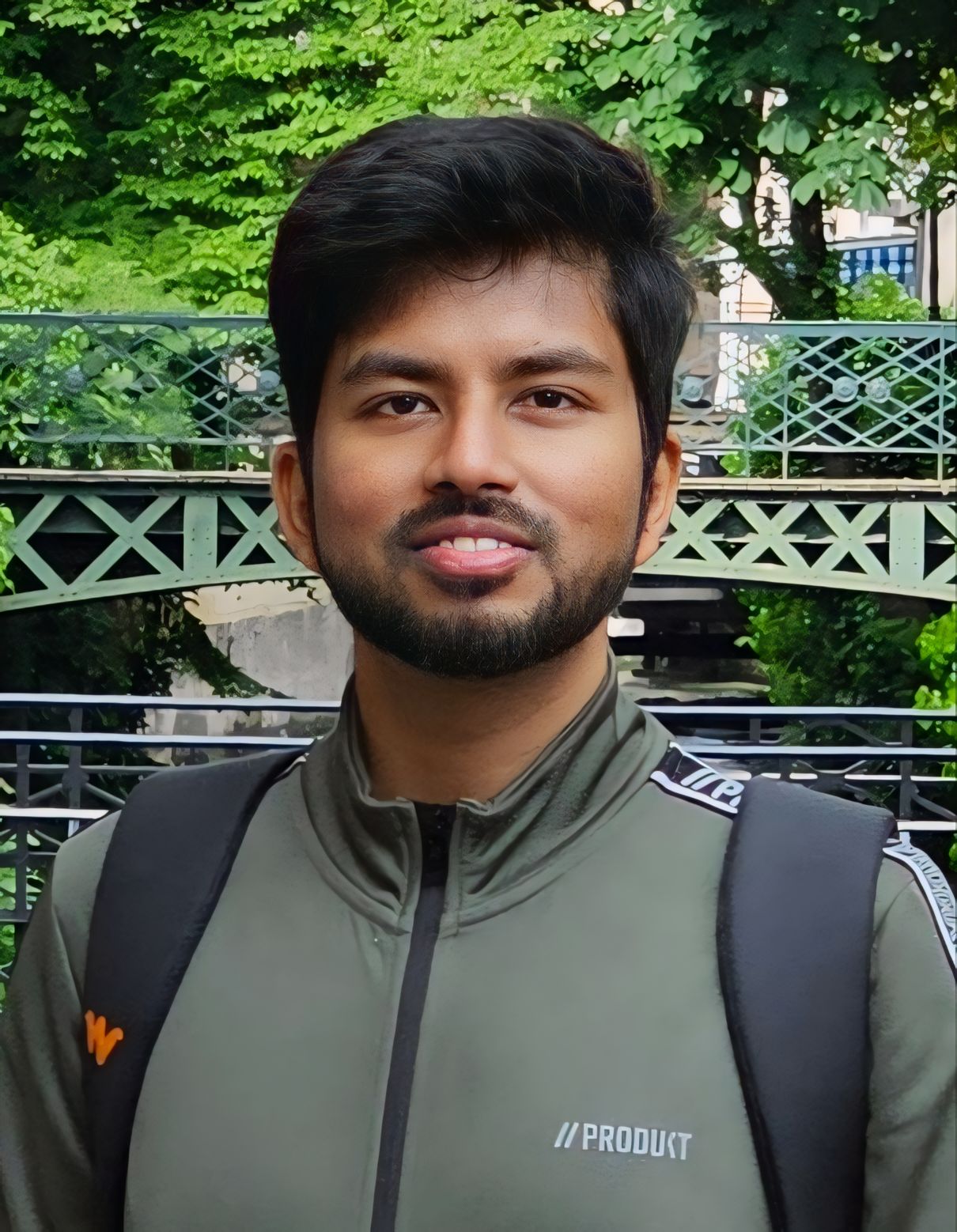}}

\begin{document}
\vspace*{4cm}
\title{Higgs and precision physics at CMS}

\author{ Spandan Mondal on behalf of the CMS Collaboration}

\address{Department of Physics, Brown University, 182 Hope Street,\\
Providence, RI 02912, USA}

\maketitle\abstracts{
Since the discovery of the Higgs boson in 2012, significant progress has been made in measuring several properties related to the Higgs boson. The large dataset available now facilitates precise measurements of the Higgs boson mass, natural width, couplings, production cross sections, and even differential and fiducial cross sections. The latest precision measurements performed by the CMS experiment in the Higgs sector are presented in this note.}

\section{Introduction}

The discovery of the Higgs boson~\cite{ATLASHiggs,CMSHiggs1,CMSHiggs2} in 2012 by the ATLAS~\cite{ATLAS} and CMS~\cite{CMS} Collaborations was a significant milestone in the field of high energy physics and a big success of the Large Hadron Collider (LHC)~\cite{LHC} physics programme. For the Higgs boson discovery, the CMS experiment made use of about 10.4 fb$^{-1}$ proton-proton (pp) collision data collected at center-of-mass energies of 7 and 8 TeV. 
Since then, the LHC has delivered nearly 20 times more data. The Run-1 (2010--2012) and Run-2 (2015--2018) of the LHC delivered 25 fb$^{-1}$ and 138 fb$^{-1}$ of pp collision data, respectively. The ongoing Run-3 has delivered about 60  fb$^{-1}$ of additional data between 2022 and 2023. The large dataset available now facilitates precise measurement of the mass of the Higgs boson, a free parameter of the standard model (SM) of particle physics which characterises, among other things, the Higgs boson production cross sections and branching fractions, and the shape of the Higgs potential. The natural width of the Higgs boson, and its coupling strength to other particles, can also be constrained with higher accuracy; any deviation from their predicted values would be a strong indication of physics beyond the SM. Finally, the large dataset allows us to perform fiducial and differential measurements to probe the Higgs kinematics in a model-independent way and search for deviations from predictions.

\section{Higgs boson mass}

The measurement of the Higgs boson mass was performed with the full Run-2 data (138 fb$^{-1}$) in the H $\rightarrow$ ZZ $\rightarrow4\ell$ channel~\cite{mass}. Events with 4 prompt and isolated leptons were selected and events were classified into 4 categories depending on the flavour of the leptons (4$\mu$, 4e, 2e2$\mu$, and 2$\mu$2e). Two Z boson candidates were reconstructed per event: Z$_1$ was reconstructed using two same-flavour opposite-sign leptons with invariant mass closest to the Z boson mass (on-shell), and Z$_2$ was reconstructed from the remaining pair (off-shell). Improvements in the momentum resolution of leptons were achieved using the following analysis techniques:
\begin{itemize}
\item The transverse momenta ($\pt$) of the leptons forming Z$_1$ were re-evaluated by using a mass constraint on the intermediate on-shell Z boson (Z$_1$).
\item The muon $\pt$ was further corrected by constraining the four lepton tracks to originate from a common vertex compatible with the beam spot; this improves the resolution in the 4-lepton invariant mass ($m_{4\ell}$) by 3--8
\item The per-event uncertainty in $m_{4\ell}$, denoted by $\delta_{4\ell}$, is evaluated by propagating per-lepton momentum uncertainty; then the events are further categorised into 9 categories based on the calculated value of $\delta_{4\ell}/m_{4\ell}$.
\end{itemize}
This results in a measured mass of $m_{\mathrm{H}}^{(\mathrm{H}\rightarrow4\ell,\ \mathrm{Run}\hbox{-}2)} = 125.04\pm0.12\ \mathrm{GeV},$
corresponding to a relative uncertainty of 0.10\%. Of this, the statistical uncertainty is about 0.11 GeV, which makes this measurement statistically limited. The dominant sources of systematics are the muon ($\sim$30 MeV) and electron ($\sim$40 MeV) momentum scales. This measurement combined with the Run-1 measurement of $m_{\mathrm{H}}^{(\mathrm{H}\rightarrow4\ell,\ \mathrm{Run}\hbox{-}1)} = 125.60^{+0.46}_{-0.45}\ \mathrm{GeV}$ yields a measurement of 
\begin{equation}
m_{\mathrm{H}}^{(\mathrm{H}\rightarrow4\ell,\ \mathrm{Run}\hbox{-}1+\mathrm{Run}\hbox{-}2)} = 125.08\pm0.12\ \mathrm{GeV},
\end{equation}
which is the most precise single-channel measurement of the Higgs boson mass. The mass measurements in different channels are shown in Fig.~\ref{fig:fig} (left).

These may be compared with the previous results obtained from 2016-only H $\rightarrow\gamma\gamma$~\cite{massgg}, 2016-only H $\rightarrow$ ZZ~\cite{massZZ}, 2016-only H $\rightarrow\gamma\gamma$/ZZ combination, and Run-1~\cite{massRun1} + 2016 H $\rightarrow\gamma\gamma$/ZZ combination, which corresponded to uncertainties of 0.21\%, 0.17\%, 0.13\%, and 0.11\%, respectively.

\section{Higgs boson width}

The Higgs boson width ($\Gamma_{\mathrm{H}}$) is constrained directly in the Run-2 H $\rightarrow4\ell$ analysis~\cite{mass} by modeling the on-shell (105 $<m_\mathrm{H}<$ 140 GeV) Higgs boson mass distribution with a Breit-Wigner distribution (convoluted with DSCB and Landau functions used to model the signal peak), to include $\Gamma_{\mathrm{H}}$ as an additional parameter in the fit to data. An upper limit of 60 MeV is obtained for $\Gamma_{\mathrm{H}}$ at the 68\% confidence level (CL) in the fit to data.

However, a stronger constraint of the Higgs boson width can be obtained by simultaneously fitting events from the Higgs on-shell and off-shell regions~\cite{mass}. This is a consequence of the off-shell production cross section having an additional dependency on $\Gamma_{\mathrm{H}}$, that the on-shell production cross section lacks:
\begin{equation}
\sigma_{\mathrm{vv\rightarrow H\rightarrow 4\ell}}^{\mathrm{on\hbox{-}shell}} \propto \mu_{\mathrm{vvH}},\  \sigma_{\mathrm{vv\rightarrow H\rightarrow 4\ell}}^{\mathrm{off\hbox{-}shell}} \propto \mu_{\mathrm{vvH}} \Gamma_{\mathrm{H}},
\end{equation}
where $\mu_{\mathrm{vvH}}$ represents the modifier of the signal strength. The Higgs boson width is measured to be $\Gamma_{\mathrm{H}}^{(\mathrm{H}\rightarrow4\ell)} = 2.9^{+2.3}_{-1.7}$ MeV at the 68\% CL, using the on-shell and off-shell regions of the Run-2 H $\rightarrow4\ell$ analysis. This measurement is consistent with the SM prediction of 4.1 MeV.

\begin{figure}

\begin{minipage}{0.46\linewidth}
\centerline{\includegraphics[width=1\linewidth]{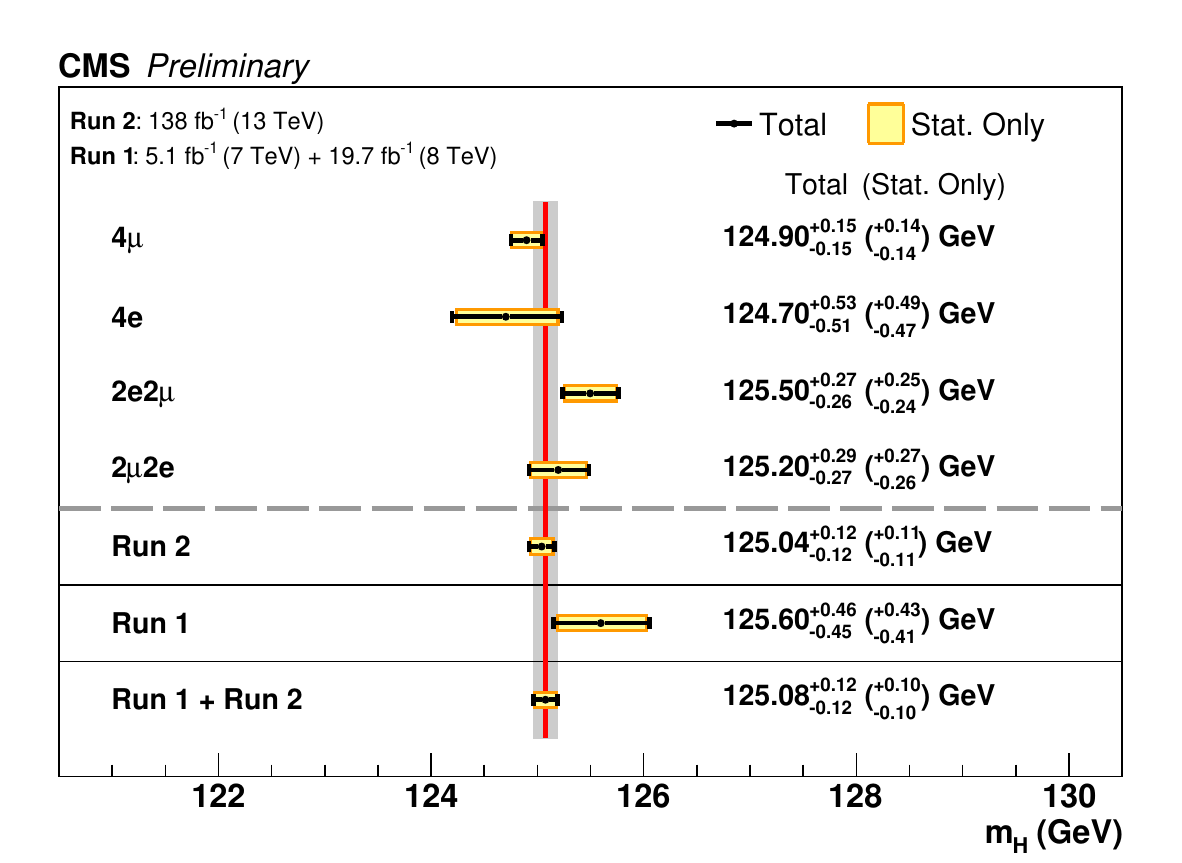}}
\end{minipage}
\hfill
\begin{minipage}{0.35\linewidth}
\centerline{\includegraphics[width=1\linewidth]{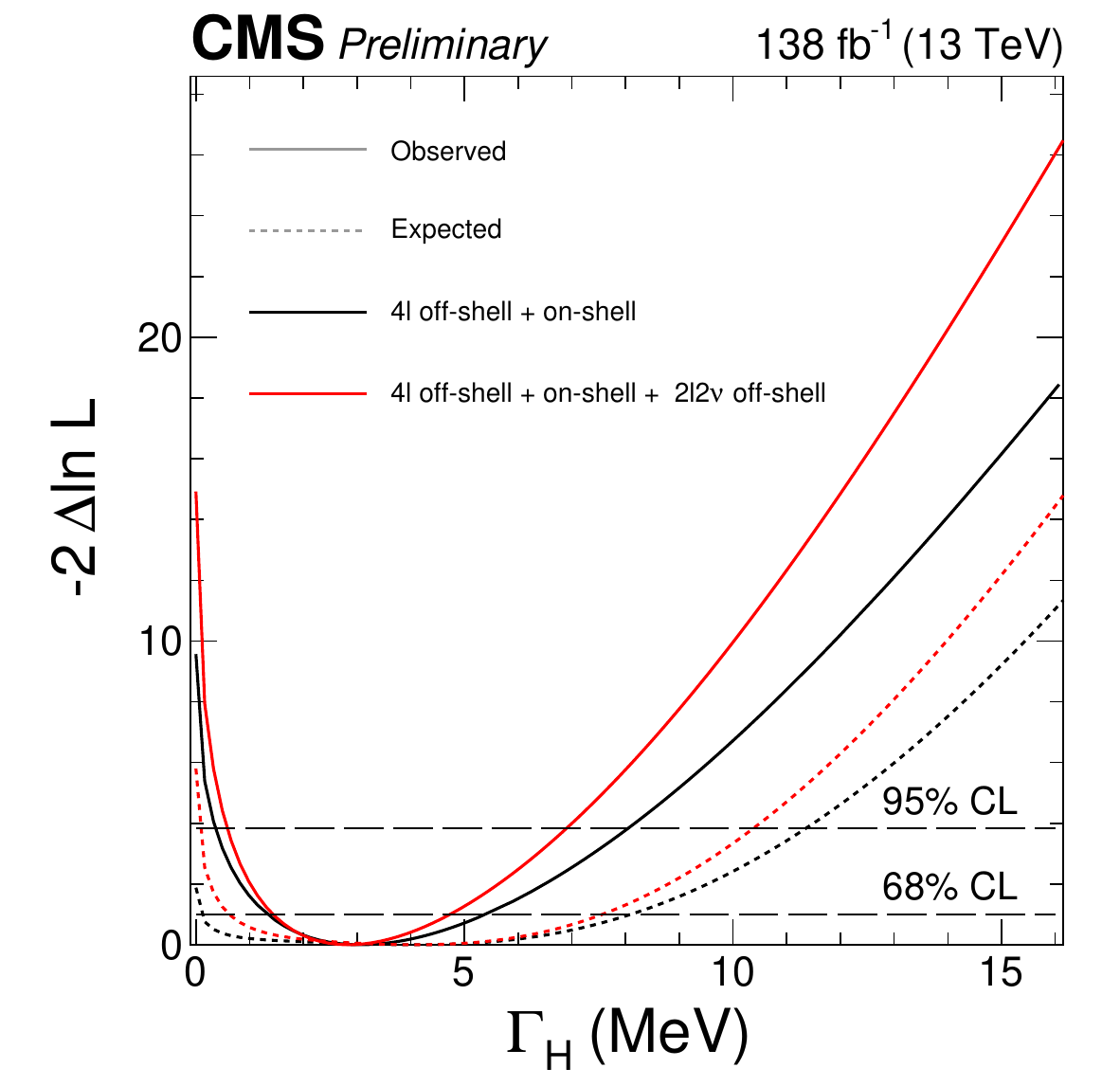}}
\end{minipage}
\caption[]{Summary of the observed CMS H boson mass measurements using the $4\ell$ final state (left) and the observed (solid) and expected (dashed) profile likelihood projection on the H boson width using the on-shell and off-shell production (right).~\cite{mass}}
\label{fig:fig}
\end{figure}

This measurement is further combined with the measurement obtained from the Run-2 off-shell H $\rightarrow ZZ \rightarrow2\ell2\nu$ analysis~\cite{offshell}, which provided the first evidence for off-shell Higgs boson contributions to the ZZ channel. The combined measurement yields a value of 
\begin{equation}
\Gamma_{\mathrm{H}}^{(\mathrm{H}\rightarrow4\ell,\ \mathrm{H}\rightarrow2\ell2\nu)} = 2.9^{+1.9}_{-1.4}\ \mathrm{MeV}
\end{equation}
at the 68\% CL. The likelihood scan for different $\Gamma_{\mathrm{H}}$ hypotheses is shown in  Fig.~\ref{fig:fig} (right).

\section{Higgs boson couplings and production modes}

The inclusive Higgs signal strength is measured to be $\mu=1.002\pm0.057$~\cite{nature}, which is about a factor of 4.5 more precise than the measurement possible at the time of the discovery. Higgs boson couplings to gauge bosons and third generation fermions have been observed individually. There has also been evidence for the Higgs boson coupling to muons~\cite{mumu}. All these measurements are in agreement with the SM predictions.

The Higgs boson coupling modifiers to the W boson ($\kappa_{\mathrm{W}}$), Z boson ($\kappa_{\mathrm{Z}}$), photon ($\kappa_{\gamma}$), gluon ($\kappa_{\mathrm{g}}$), top quark ($\kappa_{\mathrm{t}}$), and tau lepton ($\kappa_{\tau}$) have been measured with uncertainties of 10\% or lower~\cite{nature}. The couplings to b quark ($\kappa_{\mathrm{b}}$) and muons $\kappa_{\mu}$ have uncertainties of around 20\%. The measurement of $\kappa_{\mu}$ is limited by statistics and can be significantly improved with more data in the future. In terms of signal strength modifiers, the values of $\mu^{\gamma\gamma}$, $\mu^{\mathrm{ZZ}}$, $\mu^{\mathrm{WW}}$, and $\mu^{\tau\tau}$ have also been measured with uncertainties around 10\%, while $\mu^{\mathrm{bb}}$ and $\mu^{\mu\mu}$ have been measured with $\sim$20 and $\sim$45\% uncertainties, respectively.

The various productions modes of the Higgs boson have also been measured individually~\cite{nature}. The signal strength modifiers of the Higgs boson production in the gluon-gluon fusion (ggH) and Vector Boson Fusion (VBF) modes have been measured with uncertainties of $\sim$10\%. The signal modifiers for the vector boson association (ZH/WH) and top-pair associated (ttH) production modes have been measured with uncertainties of $\sim$20--30\%. The rare single-top associated production mode (tH) has not been observed yet.

\section{Higgs boson differential measurements}

The CMS experiment has performed measurements of the differential fiducial cross sections in the H $\rightarrow\gamma\gamma$~\cite{diffgg}, H $\rightarrow4\ell$~\cite{diff4l}, and H $\rightarrow\tau\tau$~\cite{difftau1,difftau2} channels. Furthermore, Simplified Template Cross Section (STXS)~\cite{STXS} measurements were performed by categorising events into simplified kinematic regions in the H $\rightarrow\gamma\gamma$~\cite{stxsgg} and H $\rightarrow$ WW~\cite{stxsww} channels. Two more recent STXS measurements are presented as follows.

\vspace{-0.08cm}\paragraph{STXS measurements in the VH(H $\rightarrow$ bb) channel  }
STXS measurements of in the H $\rightarrow$ bb decay channel and the ZH/WH production mode were performed using full Run-2 data~\cite{stxsvh}. Events were classified into three leptonic channels (0L, 1L, 2L) based on the number of reconstructed charged leptons in the event. Two Lorentz boost regimes were defined based on the $\pt$ of the vector boson (V). The so-called resolved regime is characterized by $\pt(\mathrm{V})<250$ GeV and Higgs decays are reconstructed using two b-tagged AK4 jets. Events in the boosted regime ($\pt(\mathrm{V})>250$) are reconstructed with 1 bb-tagged AK8 jet. The Higgs mass resolution is improved by using techniques like Final State Radiated (FSR) jet recovery, b jet energy regression, and kinematic fit (2L channel) using a constraint on the mass of the Z boson. Results are interpreted by measuring the signal strength modifier $\mu$ in exclusive STXS regions. No significant deviation from the SM prediction was observed in any of the STXS regions.

\vspace{-0.08cm}\paragraph{STXS measurements in the ttH(H $\rightarrow$ bb) channel  }
Another analysis~\cite{stxstth} performed by the CMS experiment targets the STXS cross sections in the H $\rightarrow$ bb decay channel and the ttH and tH production modes. Events are classified into 3 channels: fully hadronic (at least 7 jets, at least 4 are b-tagged), semileptonic (at least 5 jets, at least 4 are b-tagged, at least 20 GeV of missing energy), and dileptonic (at least 3 b-tagged jets, at least 40 GeV of missing energy).

The inclusive signal strength modifier is measured to be $0.33\pm0.26$, which is compatible with 1 (SM prediction) within 3 standard deviations. The measurements in individual STXS bins are also compatible with the corresponding SM predictions.

\section{Summary}

Significant progress has been made in the measurement of the properties of the Higgs boson since its discovery. The Higgs boson mass has been measured with a ~0.10\% precision, and its width has been constrained with an unprecedented precision. Higgs boson couplings to gauge bosons, top quark, tau lepton have been measured with 10\% precision. We are now in a precision era where differential measurements are possible. No deviations from the SM prediction have been observed so far. Constraints on Higgs boson decays that have not yet been observed~\cite{mumu,cc,Zg}, as well as on the Higgs boson self-coupling~\cite{self}, are expected to improve with the data still to be collected at the LHC and the HL-LHC.

\end{document}